# Mercury's geochronology revised by applying Model Production Functions to Mariner 10 data: geological implications

Matteo Massironi[1,2], Gabriele Cremonese[3], Simone Marchi[4], Elena Martellato[2], Stefano Mottola[5], Roland J. Wagner [5]

1 Dipartimento di Geoscienze, Università di Padova, via Giotto 1, I-35137 Padova, Italy matteo.massironi@unipd.it
2 CISAS, Università di di Padova , Italy
3 INAF, Osservatorio Astronomico di Padova, Italy
4 Dipartimento di Astronomia, Università di Padova, Italy
5 German Aerospace Center (DLR), Institute of Planetary Research, Berlin, Germany

## Abstract

Model Production Function chronology uses dynamic models of the Main Belt Asteroids (MBAs) and Near Earth Objects (NEOs) to derive the impactor flux to a target body. This is converted into the crater size-frequency-distribution for a specific planetary surface, and calibrated using the radiometric ages of different regions of the Moon's surface. This new approach has been applied to the crater counts on Mariner 10 images of the highlands and of several large impact basins on Mercury. MPF estimates for the plains show younger ages than those of previous chronologies. Assuming a variable uppermost layering of the Hermean crust, the age of the Caloris interior plains may be as young as 3.59 Ga, in agreement with MESSENGER results that imply that long-term volcanism overcame contractional tectonics. The MPF chronology also suggests a variable projectile flux through time, coherent with the MBAs for ancient periods and then gradually comparable also to the NEOs.

## 1. Introduction

From the middle 1960s onwards, the cratering records from planetary surfaces has been used to obtain age determinations for geological units and processes, as well as to make inferences about the time-dependent regimes of impactor fluxes throughout the Solar System. The Lunar crater size-frequency-distribution (SFD), whose shape is assumed to have been stable over the past 4 Gyr and to correspond to the Neukum Production Function (NPF), together with radiometric chronologies from landing sites, have provided the basis for deriving the impactor flux to the Moon, which in turn is used to estimate the SFD of crater forming projectiles for the other terrestrial bodies (Neukum et al., 2001a). This methodology, hereafter referred as the NPF model, permits the transfer of the Lunar chronology to Mercury (Neukum et al., 2001b).

A recent technique, proposed by Marchi et al. (2009), is in contrast based on dynamical models that describe the formation and evolution of the asteroids in the inner Solar System, permitting direct estimates of the impactor flux to Mercury. As in the NPF method, an age determination is obtained by cross-calibration with the Lunar chronology. This procedure, the Model Production Function (MPF) method, avoids a problematic use of scaling laws, may simulate a non-constant impactor flux through time, and, since it depends on cratering physics, allows a variable crustal layering of the target body to be taken into consideration.

In this work, the chronology of the highlands and several impact basins on Mercury has been revised using the MPF method, taking into account both the Main Belt Asteroids (MBAs) and Near Earth Objects (NEOs) dynamical models. The application of the MPF to crater counts obtained from Mariner 10 data enabled us to verify whether a change in the impactor flux was implied between ca. 4.0 and 3.5 Ga, provided important insights into the uppermost crustal layering of the Hermean surface, and gave significant age constraints for the emplacement of several smooth plains (including the Caloris interior). The geological implications of such results appear fully consistent with the original framework delineated by the findings from recent MESSENGER fly-bys.

## 2. The MPF chronology

The MPF chronology is developed taking into consideration both the NEO and MBA distributions (Bottke et al., 2002, 2005), along with the impactor velocity distribution (Marchi et al., 2005), to derive the impactor flux to Mercury's surface. The flux is then converted into the expected crater-SFD per unit time per unit surface. Such conversions rely on the crater scaling law of

Holsapple and Housen (2007), which in turn depends on the target material properties (strength, density, etc.). Finally, the absolute time calibration is obtained by comparing the estimated flux on Mercury with the Lunar flux and using the Lunar chronology. The Lunar chronology is derived by relating the radiometric age of each Apollo and Luna landing site with the corresponding cumulative number of 1 km diameter craters. The relation is governed by the following expression (Eq. (11) in Marchi et al., 2009):

$$N_1 = a(e^{bt}-1) + ct$$

where $a=1.23 \times 10^{-15}$, $b=7.85$, $c=1.30 \times 10^{-3}$, $N_1$ is the number of craters per unit surface at D=1 km and $t$ is the crater retention age (Ga). Erasing of craters has been also taken into account, but only for the crater superposition effect which, however, does not consistently affect the age determination of the regions investigated in this paper.

Strom et al. (2005) observed that the cratering on the oldest terrains on Mercury is best represented by the MBA-SFD, a result which was confirmed by Marchi et al. (2009). To further test this hypothesis, we used both NEO- and MBA-SFDs to derive two distinct MPFs and then evaluate their fit with the cumulative crater count of each geological unit.

The Holsapple and Housen (2007) scaling law depends on the physical properties of the target material. Unfortunately, the upper shell of Mercury is still poorly known (Nimmo, 2002; Nimmo and Watters, 2004). Hence, at a first approximation, we were forced to consider a crustal structure similar to the Lunar one (e.g. Toksöz et al., 1972), although we have placed the crust-mantle transition at the base of the Hermean elastic lithosphere calculated on the basis of lobate scarps geometries (Watters et al., 2002). Therefore, we have assumed a 10 km-thick layer of fractured silicates (regolith, megaregolith and heavily fractured silicates) on top of a bulk silicatic crust (strength $= 2\times10^8$ dyne cm$^{-2}$; density $= 2.8$ g/cm$^3$), in turn overlaying a peridotitic mantle (strength $= 3\times10^8$ dyne cm$^{-2}$; density $= 3.3$ g/cm$^3$) which begins at a depth of 40 km. Both strength and density increase linearly in the upper fractured layer, from 0 to $2\times10^8$ dyne cm$^{-2}$ and from 2 to 2.8 g/cm$^3$, respectively. Finally, we set a sharp transition in the scaling law, from cohesive soil in the case of small impactors to hard rock for larger ones, at a projectile size of 1/20$^{th}$ of the thickness of the heavily fractured silicate layer (i.e., 0.5 km).

## 3. Comparison between MPF and NPF chronologies

The model of Marchi et al. (2009) was adopted to derive new age estimates for several Hermean terrains previously dated by Neukum et al. (2001b). For the sake of comparability, we have used the same crater counts made on Mariner 10 data used for the first analysis; in addition, new counts of the Shakespeare and Raphael basin plains have been performed. The MPF-NEO and MPF-MBA best fits for each terrain are shown in fig.1, while a synthesis of the results is reported in tab. 1.

Both ages based on MPFs are systematically younger than the NPF ages, but within error bars. The only exception is for the Hermean highlands, which classically includes both the heavily cratered terrains and the intercrater plains (Strom and Neukum, 1988) and produces an MPF-NEO age considerably older (4.18 Ga) than the MPF-MBA and NPF ages (4.06 Ga and 4.07 Ga, respectively). In both MPF chronologies, the older basins are clearly grouped in a single narrow cluster within 3.98 and 3.92 Ga and separated from the younger basins by a time gap estimated to range from ca. 50 to ca. 100 Myr. On the other hand, the NPF age estimates show a more continuous trend of basin formation and related plains infilling. This is mainly due to the NPF age of the Beethoven plains (NPF= 3.86 Ga vs. MPF-NEO = 3.79 Ga and MPF-MBA = 3.80 Ga), which fills the time gap between the formation of the plains of Caloris and that of all the other basins (tab. 1). The discrepancy is directly related to the younger ages generated by the MPF relative to the NPF data, a difference which is particularly pronounced for the Beethoven and Chekhov plains.

The MPF-MBA curves show a better fit than the MPF-NEO ones for all the terrains except for the Shakespeare plains, for which a smaller crater-size range was considered, and for the Caloris plains where the characteristic "S" shape of the MPF curve between D~20 and D~10 km does not find a correspondence in the related crater-SFD (fig. 1). This "kink" in the MPF curve may arise from the shallower crustal layer, probably composed of silicates with a fracture density decreasing with depth. The presence of this layer causes a transition in the applied Holsapple and Housen (2007) scaling law (from cohesive to hard-rocks) set at $1/20^{th}$ of the fractured layer itself (assumed to be equal to 10 km by analogy with the average thickness of this layer in the Lunar crust). Such a kink nicely reproduces several observed cumulative distributions, such as the Dostoyevskij, Beethoven and Tolstoj plain distributions. On the other hand, it shows a merely acceptable fit with the crater-SFD of the Chekhov and Puskhin plains, a reduced fit with the Haydn and Raphael plains, and is not at all consistent with the crater distributions of the highlands and of the Caloris plains.

## 4. An heterogeneous crustal layering and the younger age of the Caloris interior plains

Several crater counts made on planetary surfaces show a distinct "kink" in the shape of cumulative crater-SFD for crater diameters lower than 20 km. In some regions of the Moon (e.g. Mare Crisium and Mare Tranquillitatis), this inflection is related to the effect of dating areas characterized by geological units partially covered by younger lava flows (Neukum and Horn, 1976; Boyce et al., 1977). In these particular cases, the crater-SFD leads to a composite SFD in which smaller craters reflect the age of the younger units, whereas the larger craters reproduce the age of the older units.

The MPF curves with their characteristic S shape, strictly linked to the physical parameters of the target material included in the Holsapple and Housen (2007) scaling law, demonstrate that, if no other processes such as geological units superposition have intervened, the inflections in the crater-SFD can be related to the rheological layering of the investigated portion of the crust. In particular, the inflection at lower crater diameters is mainly due to the transition from the cohesive to the hard rock scaling law and consequently varies with the thickness of the upper heavily fractured layer of the crust. This thickness is arguably very variable around the planet, depending on i) the rheology of a specific region, in turn related to the lithological variations and layering of the crust in that location, and ii) the age of the region itself, the fractured layer being thicker in older regions.

According to this view, the Dostoevskij, Pushkin, Chekhov and Tolstoj plains, which indicate similar ages and whose MPF curves fit more or less with the crater-SFD, are probably characterized by a very similar crustal layering with the uppermost stratum composed by ca. 10 km of fractured silicates. The crater-SFD of the Haydn and Raphael basin plains show a kink at lower diameters, and do not follow the MPF curve. In these cases, an upward migration of the inflection point of the MPF curve would not give a better fit, hence a thinner uppermost layer of fractured rocks is very unlikely. This behaviour may instead result from a composite crater-SFD in which smaller craters reflect the ages of younger lava flows emplaced over older ones.

Despite its younger age, the Beethoven plain shows a perfect fit with the MPF curves, assuming an upper fractured layer of 10 km. This may indicate the poorer rheological properties of the material infilling the Beethoven basin,. In contrast, the discrepancy between the highlands crater-SFD and the MPF curve for smaller crater diameters is most likely due to stronger crater erasing affecting these oldest units.

Interestingly, the crater-SFD of the younger Caloris interior plains suggests a much thinner uppermost fractured layer (around 4 km instead of the previous assumption of 10 km). The consequent shift of the S shape in the MPF curve gives a far better fit with the crater-SFD and implies a decrease of either 140 or 240 Ma for the age of the Caloris interior plains, considering the MBA or NEO curves respectively (Fig. 1, Tab. 1). However, the best fit for these two distributions are very similar in this case. Whatever the dynamical model used to derive the MPF curve, the final result implies an increase of the time gap with the emplacement of the Beethoven plains, for which substantial variation of the inflection point is not required. Hence, according to this new chronology (Fig. 2), the Caloris basin infilling is placed distinctly after the Late Heavy Bombardment Event (LHB, Tera et al., 1974), which, according to the radiometric age determinations of the Lunar impact melts from Apollo and Lunar meteorites, ended at around 3.8 Ga (e.g. Stöffler and Ryder, 2001).

Regarding the comparison between the degree of fit of the MPF-NEO and MPF-MBA curves, it is noteworthy that for the highlands the MBA curve shows a value of $\chi^2$ substantially smaller than that for the NEO. For all the older basins, the MBA $\chi^2$ is considerably better than for the NEO and, finally, for the younger Caloris basin the two curves show virtually the same degree of fit. Hence, the MPF-MBA is in general more suitable than MPF-NEO for the Hermean units considered in this study. The only outlier on this general trend is the Shakespeare basin for which a fitting comparison with the other basins is not reliable because the small size-range of the craters considered does not permit a check of the fit of the "S shaped" feature of the MPF curves (Fig. 1).

The variable fit of the MPF-NEO and MPF-MBA with the crater-SFD may however indicate a change in the shape of the impactor flux through time, which is more coherent with MBAs for ancient periods and gradually consistent also with NEO at the time of the Caloris Basin plains emplacement. This is in general agreement with the Strom et al. (2005) statement that the MBA coherent crater-SFD is the expression of the LHB.

## 5. The MPF models versus the new MESSENGER geological findings

Until the two first MESSENGER fly-bys in 2008 the presence of volcanism on Mercury was strongly questioned because of the lack of clear morphological evidences of volcanism and the supposedly monotonous composition of the Hermean surface, whose overall reflectance recalls the Lunar highlands (e.g. Blewett et al., 2002). In contrast, the MESSENGER data have recently provided both unambiguous morphological demonstration of the volcanic origin of most of the plains (Head et al., 2009) and clear evidence of the inhomogeneous spectral character of the surface of Mercury (Denevi et al., 2009). In addition, the newly recognized heterogeneity of the Hermean surface, attributed to differing Fe and Ti oxide contents, allowed Denevi et al. (2009) to hypothesize a layered crust made up of overlying volcanic units. The same authors speculate that also that the heavily cratered terrains may have been generated by primordial effusions. Consequently, Mercury may lack a primary crust developed from plagioclase flotation over a primordial magmatic ocean, such as is thought to have generated the Lunar highlands. Denevi et al.'s view has the non-trivial consequence that the heavily cratered terrains should be much younger than the Lunar highlands (up to 4.5 Ga in age, e.g. Stöffler and Ryder, 2001), and most probably yield ages within, or not much before, the LHB event. In other words, a substantially pre-Tolstojan age is not expected for the Hermean highlands. This is in favour of the MPF-MBA dating of the oldest units, because the MPF-MBA age of the highlands is younger that the equivalent MPF-NEO age (4.06-4.07 vs. 4.18 Ga).

Another important finding provided by the analysis of the MESSENGER data is that long-lasting volcanic activity may have characterized the Hermean crustal evolution. In particular, the crater density of the plains suggests that the volcanic activity on Mercury was sustained for a long time after the LHB; this is specifically proven by the interior and exterior Caloris plains showing a much lower density and a different crater-SFD with respect to the basin rim (Fassett et al. 2009). In addition, if the Raditladi plains are in part the result of effusive eruptions, their young age (possibly up to 1 Ga; Strom et al., 2008) may support even a very recent Hermean volcanic activity. Considering the overall contractional tectonics typical of Mercury, it is expected that the magmatic activity would have been widespread and concentrated in time during, and subsequent to, the LHB event, when intense crustal fracturing by dynamic loading should have eased effusions and a high frequency of large impacts coupled with a likely high geothermal gradient may have even caused impact induced volcanism. Afterwards the contractional tectonics would gradually have impeded volcanism by obstructing the effusion conduits opened by the impact events (Strom et al., 1975).

This view is in agreement with the recent observation that lobate scarps developed both during and after smooth plains emplacement (Watters et al., 2009). Therefore, the evolution of the planet was dominated by the conditions of equilibrium between magmatism and tectonics that were established through time depending on the thermal state of the planet (Wilson and Head, 2008). The new MESSENGER findings have shown that magmatism may have overcome contractional tectonics for a long time in places where the crust was likely to have been pervasively intersected by open effusion conduits.

Our results are consistent with widespread plain formation during the LHB event, as demonstrated by the age of most of the basin infillings. In addition, the post-LHB age of the Caloris basin, obtained through the MPF on the assumption of a thinner uppermost fractured layer, can be justified assuming a long-lasting and continuous magma infilling of the basin after the impact. This explanation is strongly consistent with the sustained and long-lasting magmatism on Mercury suggested by the MESSENGER data (Strom et al., 2008; Fassett et al. 2009) .

The time gap between the formation of older basins and the Beethoven plains, which is still inside the LHB event, and between the latter and the Caloris interior plains is more likely to be related to incomplete sampling through time of the dated geological units than to a truly episodic volcanism. However, it is very likely that the post-LHB lava emplacement would have been concentrated in locations characterized by pervasive fracturing. Hence, the uneven distribution of the ages of plains, mainly concentrated between 3.91 and 3.97 Ga, may be an indirect effect of the change in the volcanic style, which was nearly global during the LHB and focussed inside few specific basins in more recent times.

## 6. Conclusions

The geochronology of Mercury has been revised by obtaining MPF ages for several geological units on Mercury, using crater counts on Mariner 10 data. The study demonstrates that the MPF model can be a powerful tool in i) calculating surface ages considering possible influx variations through time, ii) quantitatively estimating the uppermost layering of the crust in different places of the planet (thinner top fractured layer for younger terrains), iii) making comparative inferences about the rheological properties of the materials that constitute different regions with similar ages, iv) recognizing units with composite crater-SFDs arising from superimposed lava flows. The final result is shown in fig. 2, where, besides the NPF ages, the MPF chronology is reported, assuming an MBA flux and a thinner uppermost layer of fractured silicates for the Caloris basin. The major outcome of this chronology is a younger age for all the plains, and a much younger age for the Caloris interior plains (3.59 Ga). This is in agreement with recent MESSENGER findings supporting a thermal state of the planet which could sustain volcanism over contractional tectonics for a long time.

The variable fit of the MPF-NEO and MPF-MBA with the considered crater-SFDs suggests a non-constant projectile flux, more coherent with the MBAs for ancient periods and gradually becoming consistent also with the NEOs at the time of the emplacement of the Caloris interior plains.

**Acknowledgment**

We thank an anonymous reviewer for the insightful suggestions and Simon Crowhurst for the revision of the English text.

**References**

Blewett, D. T., B. R. Hawke, and P. G. Lucey (2002), Lunar pure anorthosite as a spectral analog for Mercury, *Meteor. & Planet. Sci., 37*, 1245-1254.

Bottke, W.F., A. Morbidelli, R. Jedicke, J.M. Petit, H.F. Levison, P.Michel and T.S. Metcalfe (2002), Debiased Orbital and Absolute Magnitude Distribution of the Near-Earth Objects, *Icarus, 156,* 399-433.

Bottke, W.F., D.D. Durda, D. Nesvorný, R. Jedicke, A. Morbidelli, D. Vokrouhlicky, and H.F. Levison (2005), The fossilized size distribution of the main asteroid belt, *Icarus, 175*, 111-140.

Denevi, B. W., et al. (2009), The Evolution of Mercury's Crust: A Global Perspective from MESSENGER, *Science, 324*, 613-618.

Fassett, C.I., J.W. Head, D.T. Blewett, C.R. Chapman, J.L. Dickson, S.L. Murchie, S.C. Solomon, and T.R. Watters (2009), Caloris impact basin: Exterior geomorphology, stratigraphy, morphometry, radial sculpture and smooth plains deposits. *Earth and Planet. Sci. Lett., 285*, 297-308.

Head, J. W., et al. (2009), Volcanism on Mercury: the first MESSENGER flyby for extrusive and explosive activity and the volcanic origin of plains. *Earth and Planet. Sci. Lett., 285*, 227-242.

Holsapple, K.A. and K.R. Housen (2007), A crater and its ejecta: A interpretation of Deep Impact, *Icarus, 187,* 345-356.

Marchi, S., A. Morbidelli, and G. Cremonese (2005), Flux of meteoroid impacts on Mercury, *A&A, 431*, 1123-1127.

Marchi, S., S. Mottola, G. Cremonese, M. Massironi, and E. Martallato (2009), A new chronology for the Moon and Mercury, *AJ, 137*, 4936-4948.

Neukum, G. and P. Horn (1976), Effects of lava flows on Lunar crater population, *The Moon, 15*, 205-222.

Neukum, G., B.A Ivanov and W.K. Hartmann (2001a). Cratering Records in the Inner Solar system in relation to the Lunar reference system. *Space Sci. Rev., 96,* 55-86.

Neukum, G., J. Oberst, H. Hoffmann, R. Wagner,. and B.A. Ivanov (2001b), Geologic Evolution and cratering History of Mercury. *Planet. Space Sci., 49,* 1507-1521.

Nimmo, F. (2002), Constraining the crustal thickness on Mercury from viscous topographic relaxation, *Geophys. Res. Lett., 29*, 7-1.

Nimmo, F. and T.R. Watters (2004), Depth of faulting on Mercury: Implications for the heat flux and crustal and effective elastic thickness, *Geophys. Res. Lett., 31*, L02701.

Stöffler, D., and G. Ryder (2001), Stratigraphy and Isotope Ages of Lunar Geologic Units: Chronological Standard for the Inner Solar System, *Space Sci. Rev., 96*, 9-54.

Strom, R. G., N. J. Trask, and J. E. Guest (1975), Tectonism and volcanism on Mercury, *J. Geophys. Res., 80,* 2478-2507.

Strom, R.G., R. Malhotra, T. Ito, F. Yoshida, and D.A. David (2005), The origin of planetary impactors in the inner Solar System, *Science, 309*, 1847-1850.

Strom, R. G., C. R. Chapman, W. J. Merline, S. C. Solomon, and J. W. Head (2008), Mercury Cratering Record Viewed from MESSENGER's First Flyby, *Science, 321*, 79-81.

Strom R.G, and G. Neukum (1988). The cratering record on Mercury and the origin of impacting objects, In: Vilas, F., Chapman, C.R., Matthews, M.S. (Eds.), *Mercury*. The University of Arizona Press, 336-373.

Tera, F., D.A. Papanastassiou, and G.J. Wasserburg (1974), Isotopic evidence for a terminal Lunar cataclysm, *Earth and Planet. Sci. Lett.*, 22, 1.

Toksöz M. N., et al. (1972), Lunar Crust: Structure and Composition. *Science*, 176, 1012 – 1016.

Watters, T.R., R.A. Schulz, M.S. Robinson, and A.C. Cook (2002), The mechanical and thermal structure of Mercury's early lithosphere, *Geophys. Res. Lett., 29*, 37-1.

Watters, T. R., S. C. Solomon, M. S. Robinson, J.W. Head, S. L. André, S.A. Hauck II and S. L. Murchie, (2009), The tectonics of Mercury: The view after MESSENGER's first flyby. *Earth and Planet. Sci. Letters, 285,* 283-296.

Wilson, L., and J. W. Head (2008), Volcanism on Mercury: A new model for the history of magma ascent and eruption, *Geophys. Res. Lett., 35,* L23205.

**Figure Captions**

Figure 1: MPF minimum $\chi^2$ best fit of the cumulative crater count distributions used in this paper. For each region we report the best fit for NEO and MBA populations. Cumulative crater counts are from Neukum et al. (2001b), except for Raphael and Shakespeare. For the Caloris basin we also show a best fit obtained using a thinner fractured layer. The derived ages are shown in table 1. Error bars correspond to a variation of the minimum $\chi^2$ of $\pm 20\%$.

Figure 2: Summary of the newly derived ages (left panel) in comparison to previous estimates (right panel). See text for further details.

Table 1 : Comparison between MPF-NEO, MPF-MBA and NPF ages.

| Geological Units | MPF-NEO | | MPF-MBA | | NPF |
|---|---|---|---|---|---|
| | Age | $\chi^2$ | Age | $\chi^2$ | Age |
| Highlands (heavily cratered terrain + intercrater plains) | 4.18 ± 0.02 | 638.5 | 4.06 ± 0.02 | 66.3 | 4.07 ± 0.03 |
| Chekhov | 3.97 ± 0.01 | 1.19 | 3.98 ± 0.02 | 0.70 | 4.05 ± 0.08 |
| Dostoevskij | 3.94 ± 0.01 | 0.45 | 3.94 ± 0.01 | 0.20 | 3.99 ± 0.06 |
| Pushkin | 3.94 ± 0.03 | 1.63 | 3.94 ± 0.03 | 0.89 | 3.98 ± 0.06 |
| Raphael | 3.94 ± 0.02 | 8.55 | 3.94 ± 0.02 | 4.20 | - |
| Haydn | 3.93 ± 0.02 | 3.95 | 3.92 ± 0.03 | 1.79 | 3.99 ± 0.06 |
| Tolstoj | 3.91 ± 0.02 | 0.42 | 3.92 ± 0.02 | 0.29 | 3.97 ± 0.05 |
| Shakespeare | 3.83 ± 0.01 | 3.28 | 3.87 ± 0.02 | 4.88 | - |
| Beethoven | 3.80 ± 0.02 | 0.51 | 3.79 ± 0.02 | 0.20 | 3.86 ± 0.05 |
| Caloris (interior) | | | | | |
| top layer (10 km) | 3.74 ± 0.04 | 1.72 | 3.73 ± 0.07 | 2.64 | 3.77 ± 0.06 |
| top layer (4.3 Km) | 3.50 ± 0.03 | 0.16 | 3.59 ± 0.02 | 0.10 | |

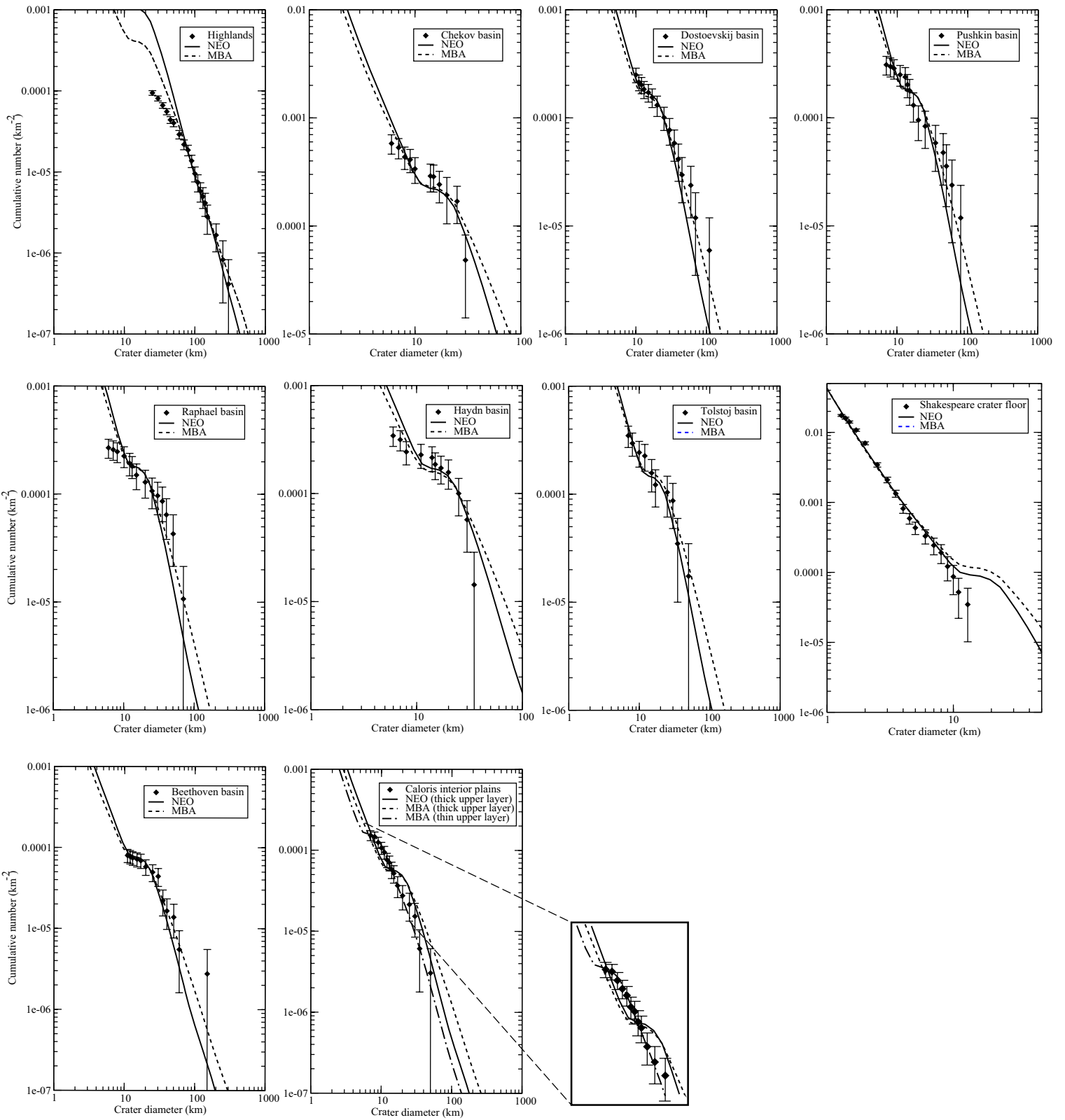

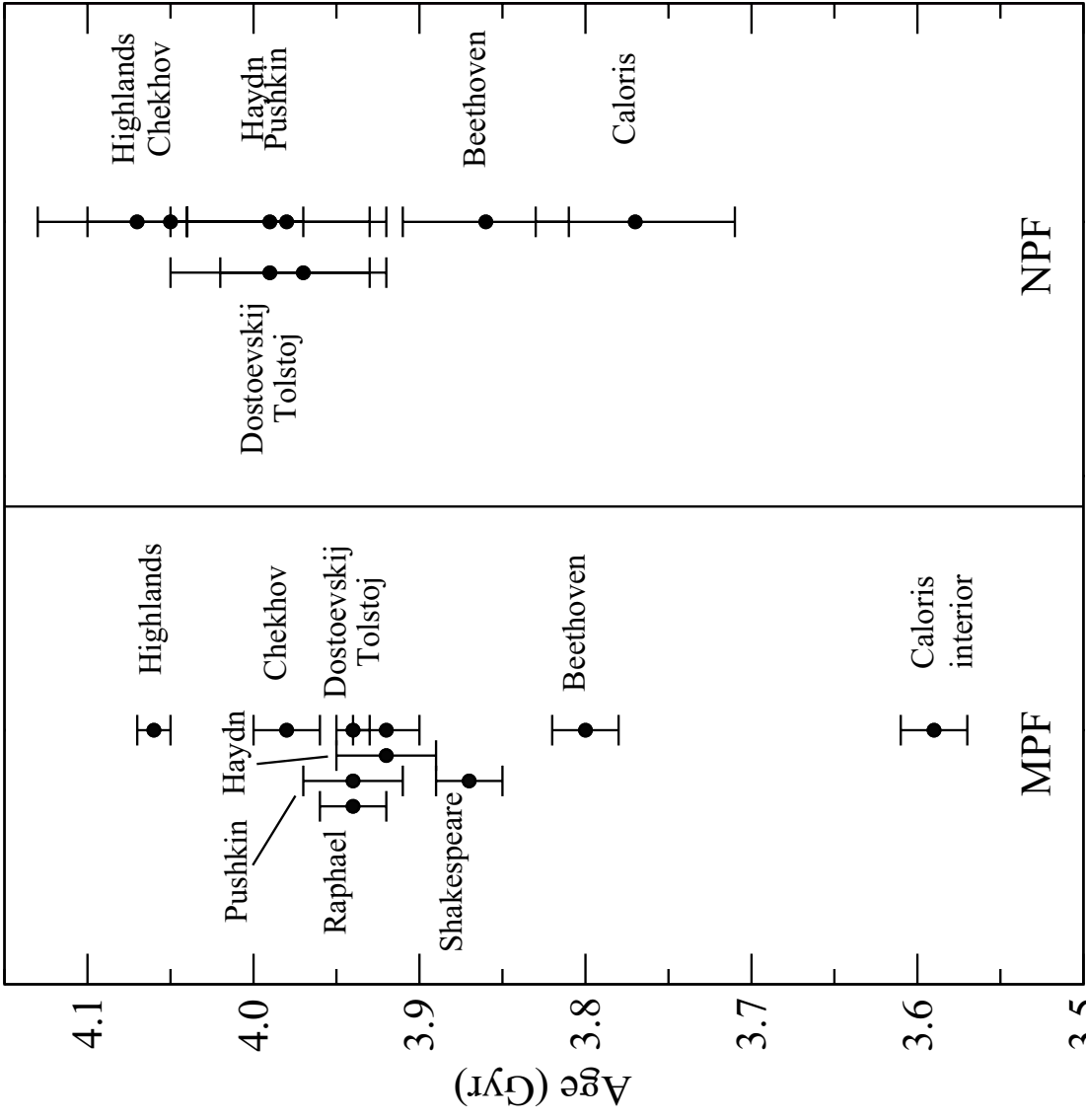